\documentclass[twoside]{ilcws10}
\usepackage[latin1]{inputenc}
\usepackage[dvips]{graphicx,epsfig,color}
\usepackage{wrapfig,rotating}
\usepackage{amssymb,amsmath,array}

\newcommand{\polRL}{e^-_R e^+_L}
\newcommand{\polLR}{e^-_L e^+_R}

\begin{document}

\title{
Higgs Recoil Mass and Higgs-Strahlung Cross-Section Study for the ILD LOI}

 \author{Hengne LI \\
\vspace{.3cm}\\
Laboratoire de l Acc\'el\'erateur Lin\'eaire (LAL),   \\
91898 Orsay Cedex, France \\
\vspace{.1cm}\\
Laboratoire de Physique Subatomique et de Cosmologie (LPSC),\\
38026 Grenoble Cedex, France\\
}

\maketitle

\begin{abstract}
This proceeding summarizes the Higgs recoil mass and Higgs-strahlung cross-section study done for the Letter of Intent (LOI) of the International Large Detector (ILD) Concept. Assuming $M_{H}=120~\rm{GeV}$, working at $\sqrt{s}=250~\rm{GeV}$ with beam parameters RDR250 and beam polarization $P(e^{-},e^{+})=(-80\%,+30\%)$, this full simulation study predicts that, the ILD detector can achieve $37~\rm{MeV}$ precision on the $M_{H}$ measurement and $3.3\%$ on the cross-section measurement from the $ZH\rightarrow \mu^{+}\mu^{-}X$ channel, while $83~\rm{MeV}$ and $4.9\%$ from the $ZH\rightarrow e^{+}e^{-}X$ channel, if we have $250~\rm{fb^{-1}}$ integrated luminosity.
\end{abstract}

\section{Introduction}
Experimental conditions at the proposed International Linear Collider~\cite{ilc} (ILC) provide an ideal environment for the precision studies of Higgs production own to the unparalleled cleanliness and well-defined initial conditions. We can use the Higgs-strahlung process ($e^{+}e^{-}\rightarrow Z^{*}\rightarrow ZH$), which is a major Higgs production mechanism at the ILC, to precisely measure the Higgs mass ($M_H$) independent of its decay modes, using the mass recoiling to the $Z$ boson, with $Z\rightarrow \mu^+\mu^-$ or $e^+e^-$. The major advantage of this recoil mass method is we do not need any information about how the Higgs decays. To calculate the $M_H$, we only need the $Z$ boson reconstructed from the lepton pair it decays with the precisely known center of mass energy ($\sqrt{s}$). We can also precisely determine the Higgs-strahlung cross-section ($\sigma$) and therefore the coupling strength at the $HZZ$ vertex, without any bias from the Higgs decay assumptions.

This proceeding summarizes the Higgs recoil mass and Higgs-strahlung cross-section study done for the Letter of Intent~\cite{ildloi} (LOI) of the International Large Detector~\cite{ild} (ILD) Concept. It is a most realistic full simulation study up to now based on the ILC design issued in the ILC Reference Design Report~\cite{rdr} (RDR) and the ILD detector concept. Thus, it has contemplated all the major effects we have thought of from both the accelerator and the detector. It assumes $M_H=120~\rm{GeV}$, the ILC works at $\sqrt{s}=250~\rm{GeV}$, and an integrated luminosity of $250~\rm{fb^{-1}}$ (equivalent to four years of data taking). The study also assumes two options of beam longitudinal polarization, $\polLR$: $(e^-:-80\%,~e^+:+30\%)$ and $\polRL$: $(e^-:+80\%,~e^+:-30\%)$, to quantitatively estimate the benefits from the beam polarization. 

The results of this study are published in the ILD LOI, with an accompany physics note~\cite{note} summarizing the methods. For the complete documentation of this study, please refer to my Ph.D thesis~\cite{thesis}. 

\section{Production}

The event generation is centrally done at SLAC~\cite{slac} using WHIZARD~\cite{whizard} v1.40, based on a detailed beam simulation using GUINEA-PIG~\cite{guineapig} with beam parameters given in ILC RDR, in which the beam energy spread are 0.28\% and 0.18\% for electron and positron beams, respectively. Both the Initial-State Radiation (ISR) and Final-State Radiation (FSR) are included in the event generation. The full simulation of the ILD detector and the reconstruction are done at DESY and KEK using software package ILCSoft~\cite{ilcsoft} v01-06. 

Table~\ref{tab:xsec} give the cross-sections of signal and major backgrounds considered in this study. We categorize the backgrounds according to the final states. For instance, the $ee$ background consists of $e^+e^-\rightarrow Z/\gamma^* \rightarrow e^+e^-$, the $\mu\nu\mu\nu$ background is mostly the $WW\rightarrow \mu\nu\mu\nu$, but also $ZZ\rightarrow\mu\mu\nu\nu$, while the $eeff$ background comes from all the possible intermediate states $e^+e^-\rightarrow Z/\gamma^*Z/\gamma*$.

\begin{table}[htb]
\centering
\begin{minipage}[b]{0.5\textwidth}
\centering
\begin{tabular}{c|cc}
Process &  \multicolumn{2}{c}{Cross-Section} \\
\cline{2-3}
	      & $\polLR$ 	& $\polRL$ \\
\hline
\boldmath{$\mu\mu X$} & \bf{11.67 fb} & \bf{7.87 fb}\\
$\mu\mu$ & 10.44 pb & 8.12 pb\\
$\tau\tau$ & 6213.22 fb  & 4850.05 fb \\
$\mu\nu\mu\nu$ & 481.68 fb  & 52.37 fb \\
$\mu\mu ff$ & 1196.79 fb  & 1130.01 fb \\
 \hline

\end{tabular}
\end{minipage}%
\begin{minipage}[b]{0.5\textwidth}
\centering
\begin{tabular}{c|cc}
Process &  \multicolumn{2}{c}{Cross-Section} \\
\cline{2-3}
	      & $\polLR$ 	& $\polRL$ \\
\hline
\boldmath{$ee X$} & \bf{12.55 fb} &  \bf{8.43 fb}\\
 \hline
$ee$ & 17.30 nb  & 17.30 nb\\
 \hline
$\tau\tau$ & 6213.22 fb  & 4814.46 fb\\
 \hline
$e\nu e\nu$ & 648.51 fb  & 107.88 fb \\
 \hline
$eeff$ & 4250.58 fb  & 4135.97 rb\\
 \hline
\end{tabular}

\end{minipage}
\caption{Cross-sections with the two options of beam polarization $\polLR$ and $\polRL$. The signal is indicated by bold face letters.}
\label{tab:xsec}
\end{table}

\section{Event Selection}

The first step in the event selection is the identification of leptonically decaying $Z$ bosons. We require the lepton tracks to be well measured by removing tracks with large uncertainties on the reconstructed momentum.The efficiency to identify a pair of leptons from the decay of a $Z$ boson is 95.4~\% for $Z\rightarrow\mu^+\mu^-$ and 98.8~\% for $Z\rightarrow e^+e^-$. 

The second step in the event selection is to suppress the backgrounds who also have a pair of leptons identified. We can suppress most of them using some common kinematic requirements.  These requirements are, the invariant mass of the lepton pair ($M_{dl}$) is around the $Z$ boson mass, a large transverse momentum ($P_{Tdl}$) of the lepton pair, a not-so-forward polar angle ($\cos{\theta}_{dl}$) distribution of the lepton pair system, the pair of leptons is not back-to-back (acoplanarity), and a restrict on the recoil mass ($M_{recoil}$) window around the Higgs mass. 

There are irreducible backgrounds like lepton pairs ($\mu\mu$ or $ee$), $WW$ and $ZZ$ after the requirements above. 

For the irreducible lepton pair background, we have found another method named $P_T$ balance. The idea is to identify the energetic ISR photons, and ask them to balance the $P_T$ of the lepton pair. We can observe a great $P_T$ balance from the lepton pair background, while not from the $ZH$ signal. This is because, the irreducibility of the lepton pair backgrounds is owen to the fact that the ISR energy loss reduces the effective $\sqrt{s}$, thus migrates them to the acceptable window of our common kinematic requirements.  Together with a veto of the photon conversions by requiring a large polar angle difference between the pair of leptons, we can totally remove the lepton pair backgrounds.

For the irreducible $WW$ and $ZZ$ backgrounds, we employ a Likelihood method for the further suppression. We use variables $M_{dl}$, $P_{Tdl}$, $\cos{\theta}_{dl}$ and acollinearity to build the templates for Likelihood calculation, and optimize the cut on the Likelihood according to the maximum of the significance. The Likelihood further suppression rejects the $WW$ and $ZZ$ backgrounds further by one half.

With this event selection scenario, we can keep an efficiency of around 55~\% with a signal over background ratio of about 4 under the recoil mass peak for the $ZH\rightarrow\mu\mu X$ channel.

\section{Fitting Methods}

After the event selection, the remaining spectrum consists of the $ZH$ signal (S) and the irreducible backgrounds (B). Now we need to build a composed model $F_M(x;~M_H, N_S)$ $=$ $N_S$ $\cdot$ $F_S(x;~M_H)$ $+$ $N_B$ $\cdot F_B(x)$, and fit it to the remaining spectrum to extract our results. In the composed model, the $M_H$ and $N_S$ are the free parameters we need to determine. $M_H$ is the Higgs mass, and the $N_S$ is for the calculation of the cross-section. 

We choose the polynomial function to describe the shape of the background. For the signal,  I have studied three functions in my thesis. They are: 

\begin{itemize}
     \item Use Gaussian to describe the Peak and adding an Exponential to complement the Tail (GPET). It is function modified from previous contributions~\cite{martin, wolfgang}. My modification makes both itself and its first derivative being continuous. 
     \item An implementation of the Kernel Estimation~\cite{ke} using Gaussian kernel. 
     \item A Physics Motivated function first developed in my thesis. The development started from Yokoya-Chen's Beamstrahlung approximation~\cite{yc1, yc2, yc3, yc4}, analytically convoluted with the ISR approximation~\cite{isr}. After a numerical convolution with the Gaussian function, I propagated it to the recoil mass. With this Physics Motivated function, and known the beam parameters, I can essentially predict the generator level $ZH$ spectrum.
\end{itemize}

\begin{figure}[hp]
\centering
\includegraphics[width=0.5\textwidth]{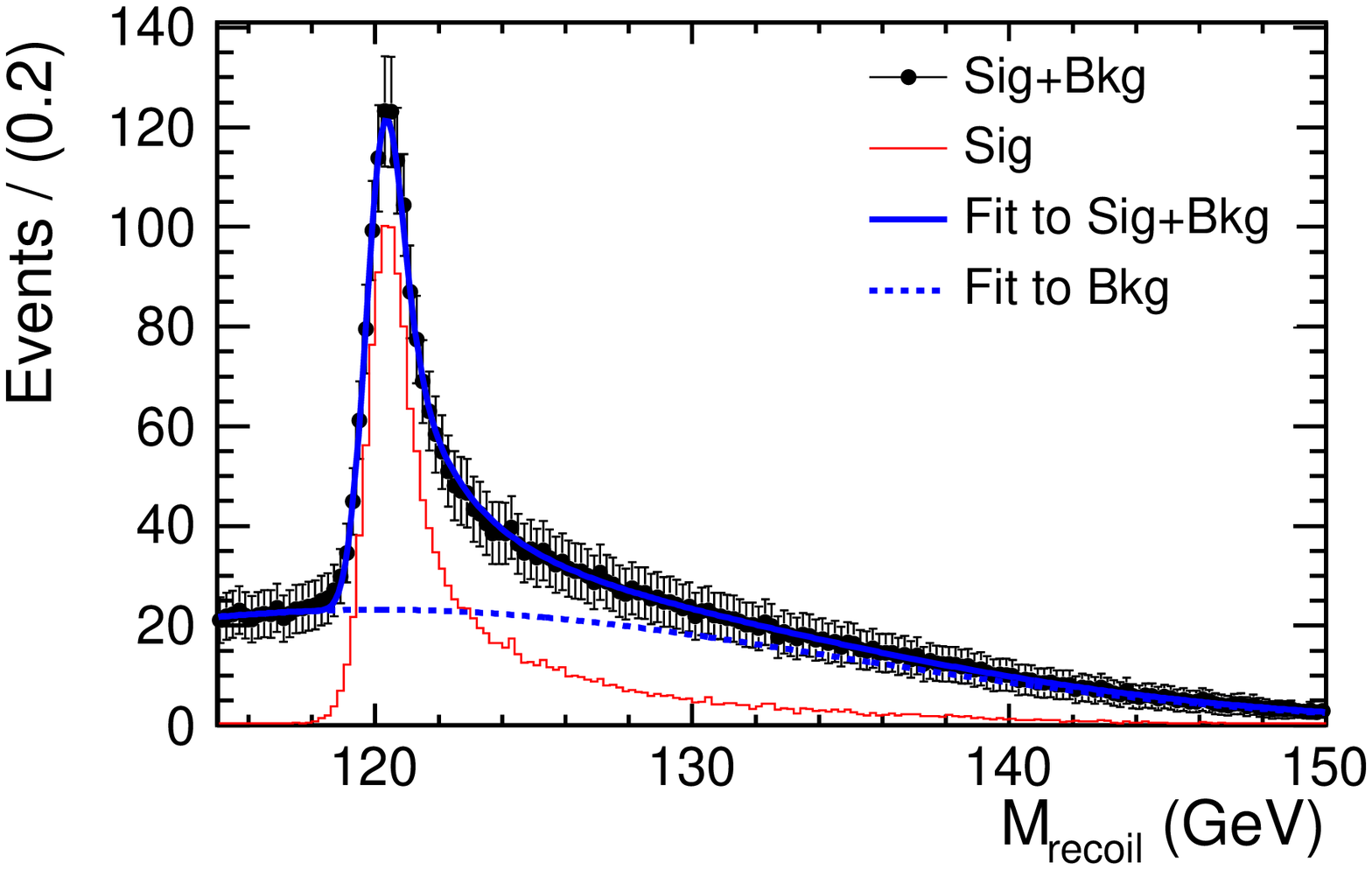}%
\includegraphics[width=0.5\textwidth]{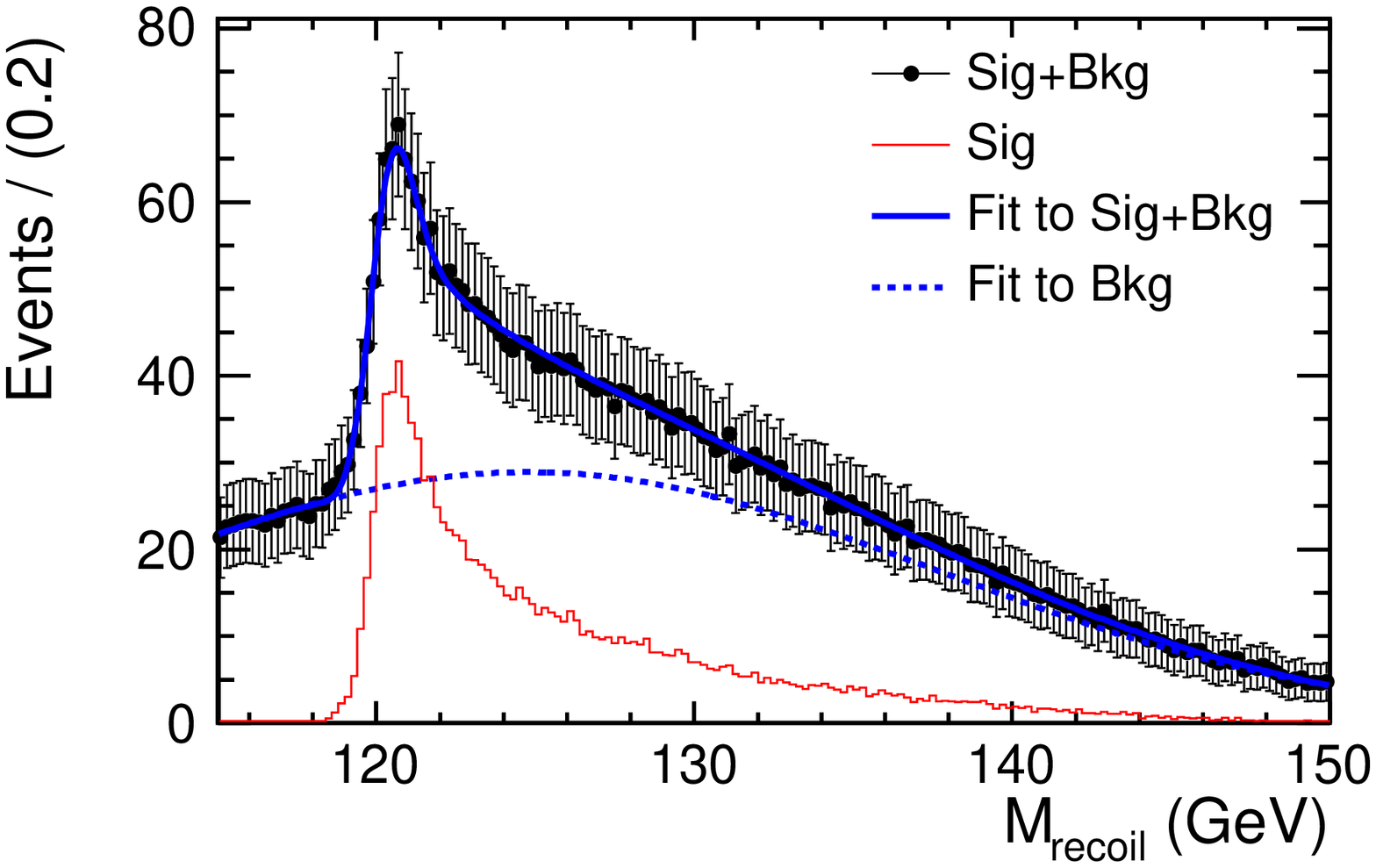}
\caption{Fits using the Physics Motivated function to describe the signal, for $ZH\rightarrow \mu\mu X$~(Left) and $eeX$~(Right) channels, taking beam polarization $\polLR$ as examples.}
\label{fig:fit0}
\end{figure}

Figure~\ref{fig:fit0} shows the fits using the Physics Motivated function to describe the signal, for both the $ZH\rightarrow \mu\mu X$ and $eeX$ channels, taking beam polarization $\polLR$ as examples. The precisions derived from the fits are 37~MeV on the $M_H$ measurement and 3.3~\% on the $\sigma$ measurement from the $\mu\mu X$ channel, while 83~MeV and 4.9~\% from the $eeX$ channel.

\section{Bremsstrahlung Recovery}

The results derived above shows the $eeX$ channel gives almost twice the errors on both the $M_H$ and the $\sigma$ measurements, than those from the $\mu\mu X$ channel. This is because the electrons suffer heavily from the Bremsstrahlung energy loss, reminding that the ILD detector has about $4~\%~X_0$ material budget before TPC. 

\begin{figure}[hp]
\centering
\includegraphics[width=0.466\textwidth]{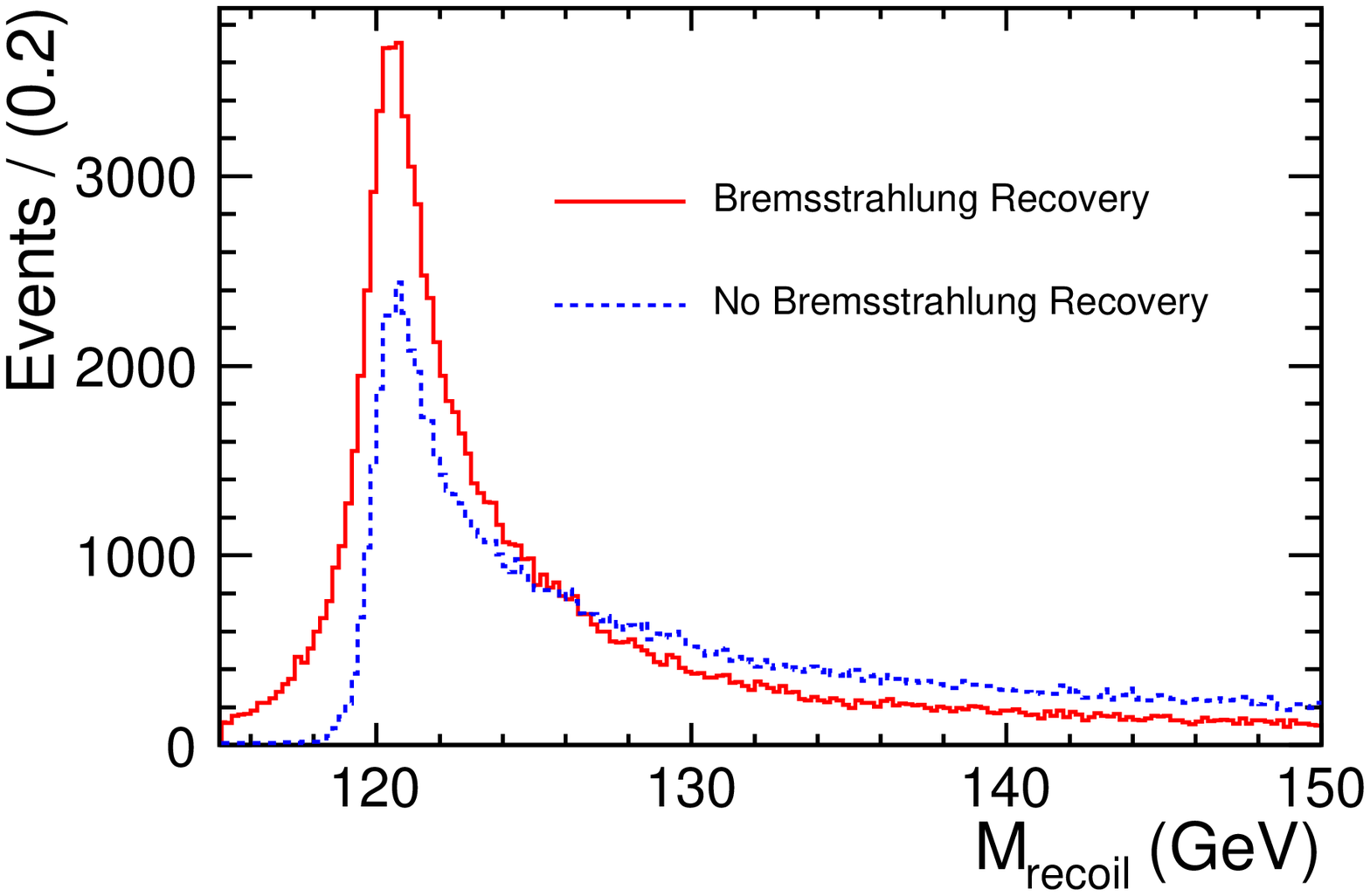}%
\includegraphics[width=0.5\textwidth]{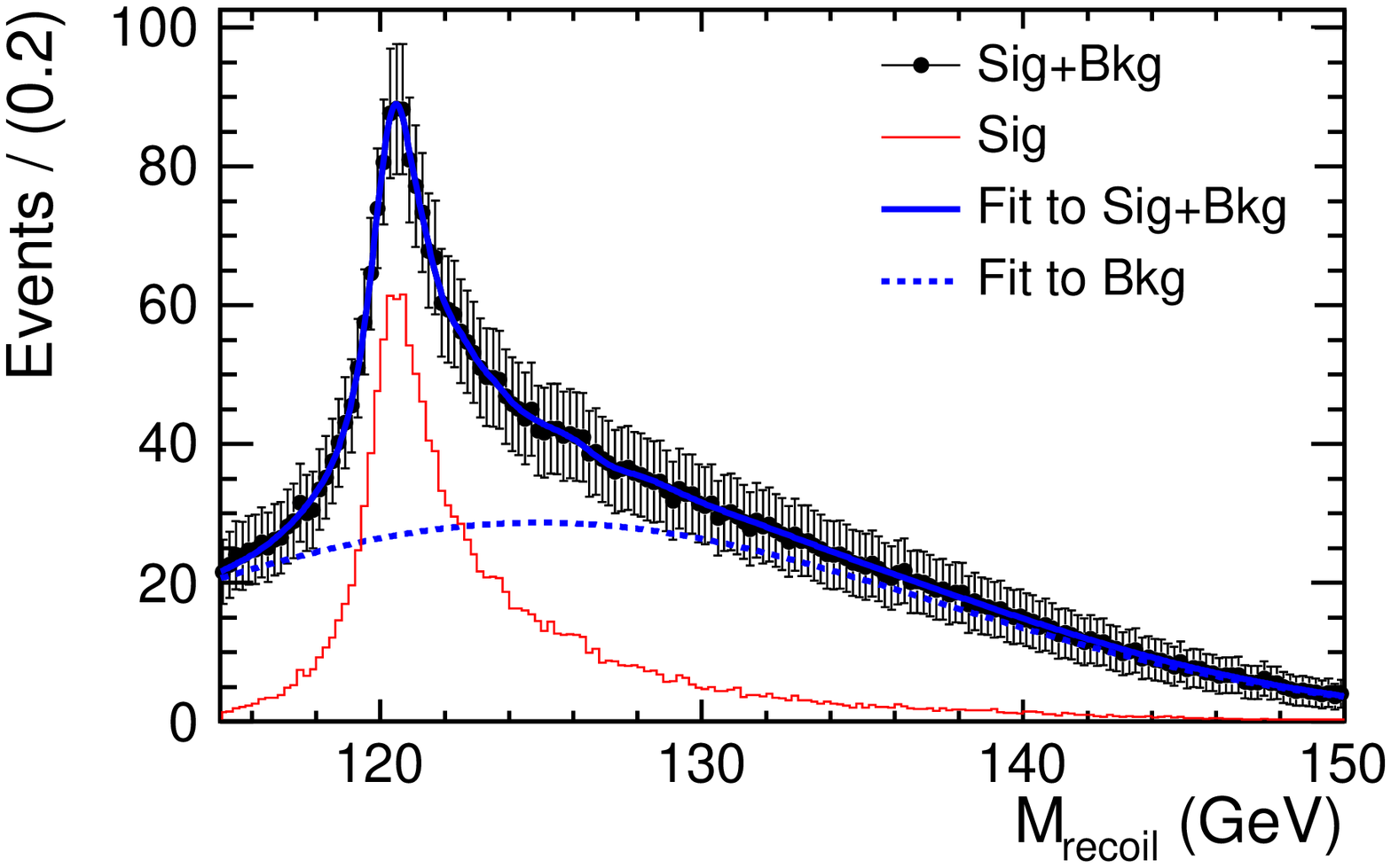}
\caption{Left: Comparison of the recoil mass distributions with and without Bremsstrahlung recovery. Right: Fit of the $eeX$ channel with Bremsstrahlung recovery corresponding to that in Figure~\ref{fig:fit0}.}
\label{fig:br}
\end{figure}

A dedicated algorithm~\cite{zfinder} is employed in this study to identify the Bremsstrahlung photons and merge them into the $Z$ boson. Figure~\ref{fig:br}~(Left) shows a comparison of the Higgs recoil mass spectra with and without Bremsstrahlung recovery. We can observe the Bremsstrahlung recovery significantly increases the statistics while degrades the mass resolution because of the worse ECAL energy resolution for low energy photon reconstruction.

Figure~\ref{fig:br}~(Right) shows the fit of the $eeX$ channel with Bremsstrahlung recovery corresponding to that in Figure~\ref{fig:fit0}. The results with Bremsstrahlung recovery are 73~MeV on the $M_H$ measurement ($\sim10~\%$ improvement), and 3.9~\% on the $\sigma$ measurement. With the Bremsstrahlung recovery, the accuracy of the $M_H$ measurement is still worse than that of the $\mu\mu X$ channel by a factor of two. However, the accuracy of the cross section measurement becomes similar to that of the $\mu\mu X$ channel, since it is less sensitive to the mass resolution.

\section{Results}

Table~\ref{tab:result} summarizes the resulting precisions of the measurements for both beam polarization $\polLR$ and $\polRL$, where those of the $eeX$ have Bremsstrahlung recovered, and the ``merged'' means the results by combining the two leptonic channels. 

\begin{table}[tb]
\centering
\begin{tabular}{|c|l|l|l|}
\hline
Pol.		&	Ch.			&	$M_H$ (GeV)		&	$\sigma$ (fb) 			\\ \hline						
		&	$\mu\mu X$	&	120.001$\pm$0.037	&	11.63$\pm$0.39(3.35\%)	\\ \cline{2-4}
$\polLR$	&	$eeX$		&	119.997$\pm$0.073	&	12.52$\pm$0.49(3.91\%)	\\ \cline{2-4}
		&	merged		&	120.006$\pm$0.033	&	12.02$\pm$0.31(2.54\%)	\\ \hline
		&	$\mu\mu X$	&	119.997$\pm$0.040	&	7.82	$\pm$0.28(3.58\%)	\\ \cline{2-4}
$\polRL$	&	$eeX$		&	120.003$\pm$0.081	&	8.41	$\pm$0.36(4.28\%)	\\ \cline{2-4}
		&	merged		&	120.005$\pm$0.035	&	8.09	$\pm$0.22(2.73\%)	\\ \hline
\end{tabular}
\caption{Resulting Higgs mass $M_H$ and cross section $\sigma$, of the $\mu\mu X$ channel and $eeX$ channel with Bremsstrahlung recovery. The ``merged'' means the results by combining the two leptonic channels. }
\label{tab:result}
\end{table}

\section{Conclusion}

The results in Table~\ref{tab:result} shows the best results of a single leptonic channel come from $ZH\rightarrow \mu\mu X$. Even including the Bremsstrahlung recovery, the $eeX$ channel gives worse results than the $\mu\mu X$ channel by a factor of two in the $M_H$ measurement, and by a factor of 1.5 in the $\sigma$ measurement. This is because of the worse energy resolution of the soft photon measurement by the ECAL, and larger backgrounds related to $eeX$ channel.  

The results with beam polarization $\polLR$ are better than those of the $\polRL$ by about 10~\%. The reasons are that, although the polarization $\polRL$ suppresses the $WW$ background, the cross sections of the Higgs-strahlung process are smaller by about 20~\% compared to that of the polarization $\polLR$. At the same time, the methods we have developed are efficient enough for the suppression of the $WW$ background. Hence the polarization $\polLR$ gives better results.

We may also conclude that there are two accelerator effects, the beam energy spread and the Beamstrahlung, have great impact on Higgs recoil mass measurement. The beam energy spread increases the width of the recoil mass peak, thus reduces the accuracy of the mass measurement. The Beamstrahlung largely reduces the effective statistics on the recoil mass peak. 

\begin{figure}[!htbp]
\centering
\includegraphics[width=0.5\textwidth]{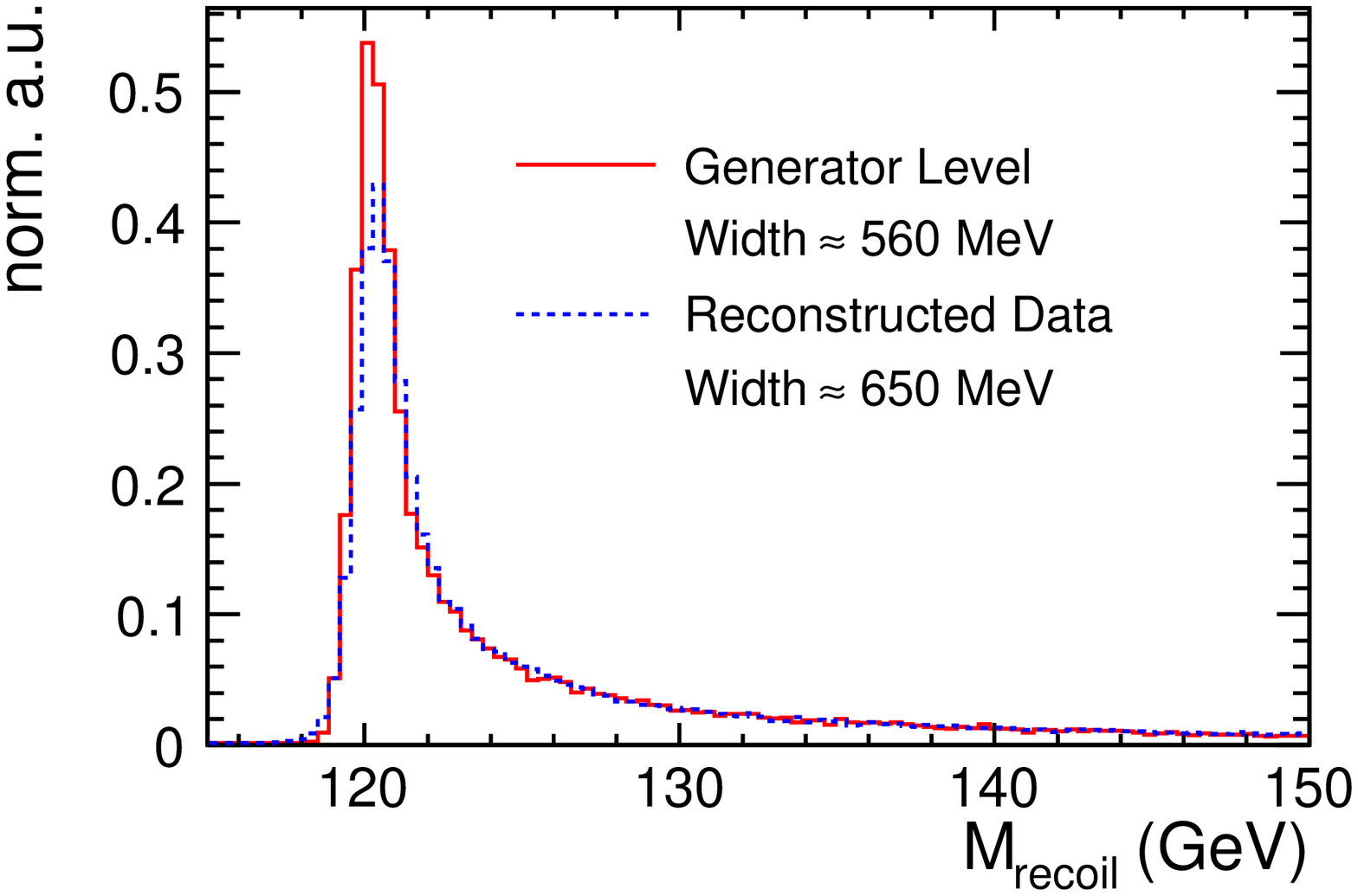}%
\includegraphics[width=0.5\textwidth]{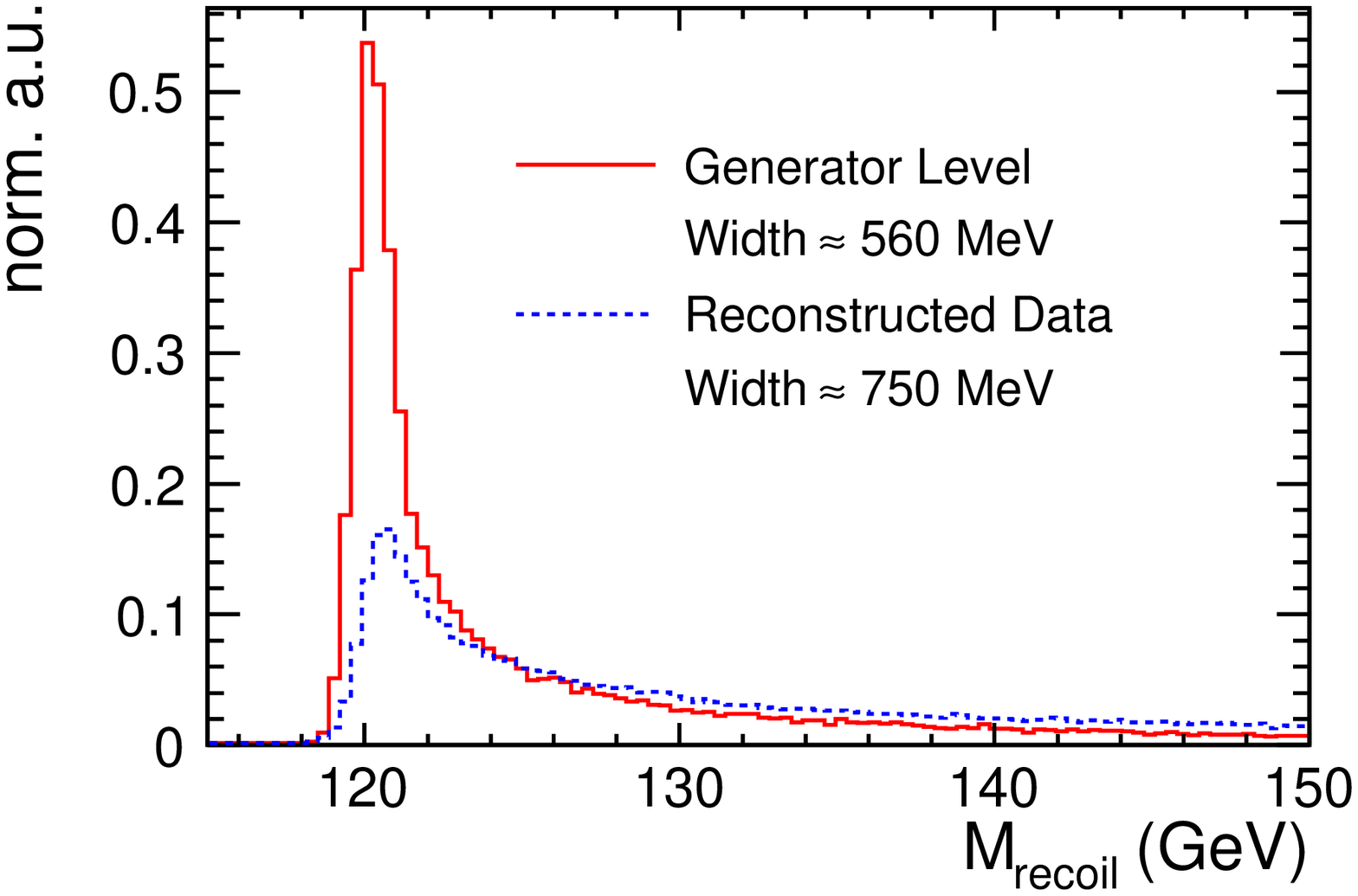}
\caption{The Higgs recoil mass distributions in the $\mu\mu X$ channel (Left) and $eeX$ channel (Right), comparison of those before and after detector simulation.} 
\label{fig:mh_gen_sim}
\end{figure}

Figure~\ref{fig:mh_gen_sim} compares the Higgs recoil mass distributions before (generator level) and after (reconstructed) full detector simulation and reconstruction for $\mu\mu X$ channel (Left) and $eeX$ channel (Right). The distribution in the generator level has the accelerator effects imposed, while that after reconstruction has the detector effects added. 
 
For the recoil mass distributions of $\mu\mu X$-channel, a fit to the left side of the maximum using a Gaussian function gives the mass resolutions to be 560 MeV in the generator level, and 650 MeV after full detector simulation. The detector response leads to a broadening of the recoil mass maximum from 560 MeV to 650 MeV. The contribution from the uncertainty of detector response is therefore estimated to be 330 MeV. This observation indicates that \emph{the dominant contribution to the observed width of the $\mu\mu X$ recoil mass distribution arises from the incoming beams rather than the response of the ILD detector}.  

\section{Outlook}

All what we have intensively studied in the past is how to reduce the statistical errors on these two measurements. A serious study of the systematic errors is apparently missing in the context. 

In the Higgs recoil mass measurement, the systematic biases appear as the difference between the measured value and the true value. Observables correlated with the Higgs recoil mass are the center of mass energy and the momenta of the pair of leptons. Imperfect knowledge of these variables could introduce systematic bias to the Higgs recoil mass measurement. To control them, a well studied reference reaction is needed.

\begin{figure}[h]
\centering
\includegraphics[width=0.5\textwidth]{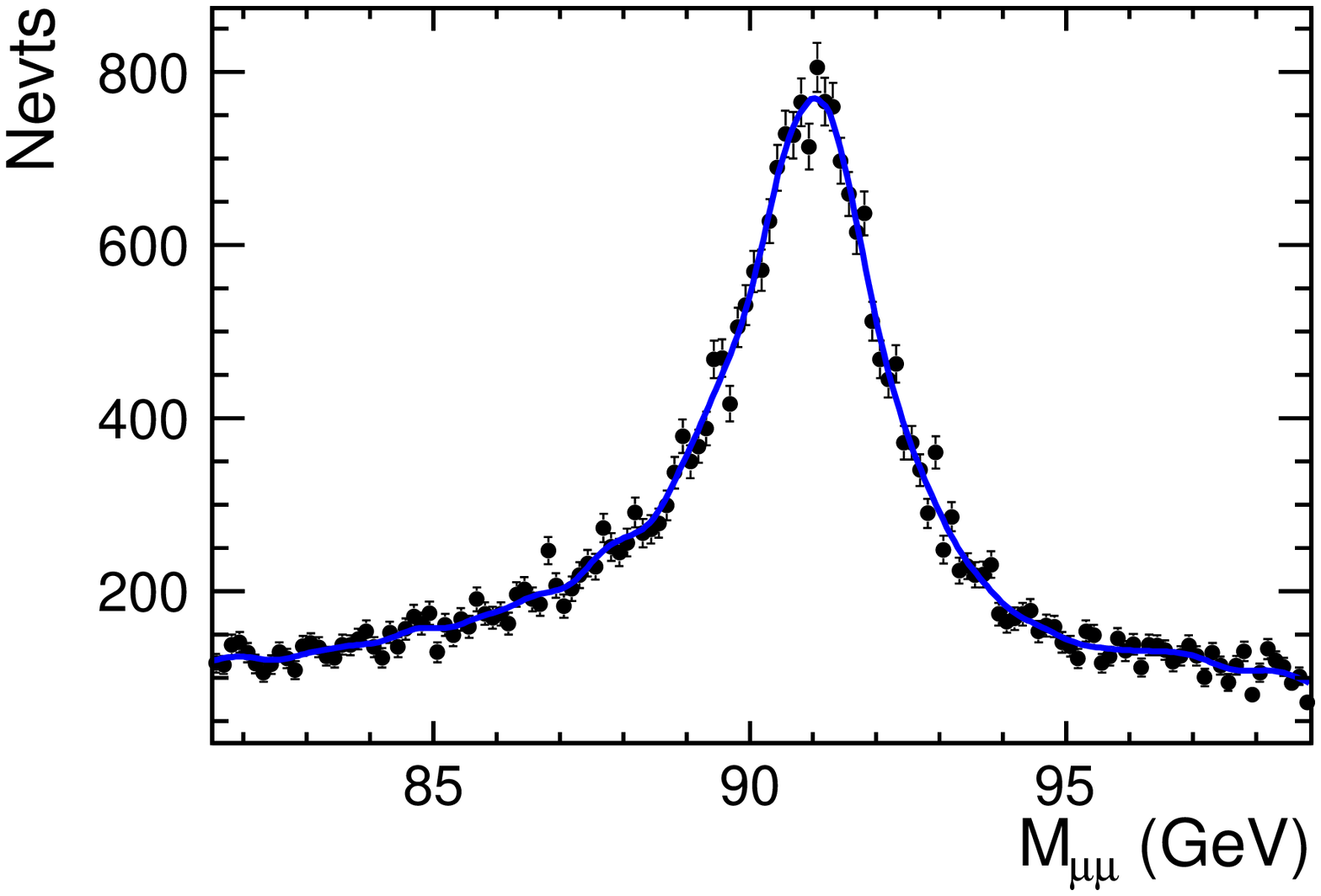}%
\includegraphics[width=0.5\textwidth]{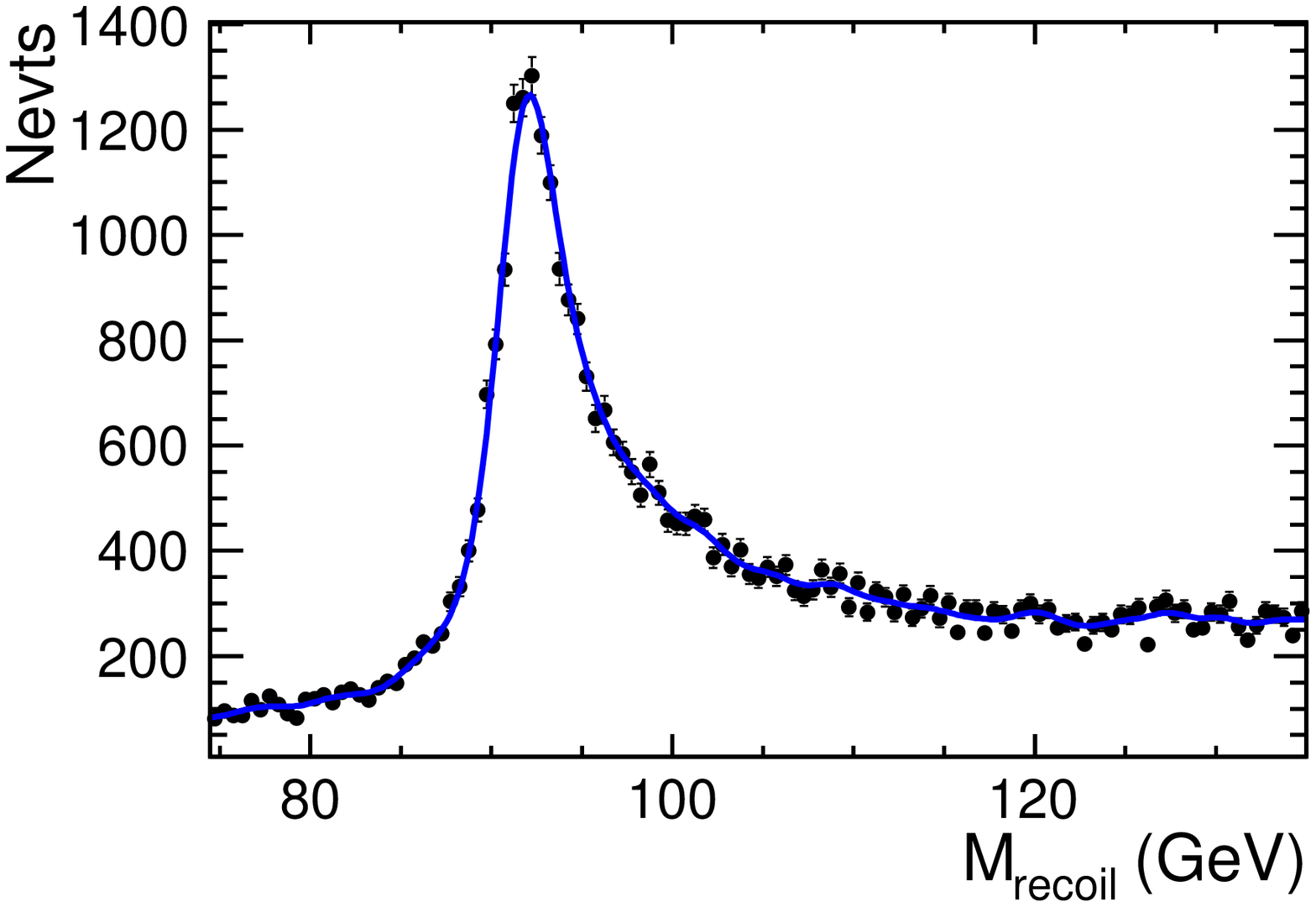}
\caption{Fit of the \emph{Invariant Mass} (Left) and \emph{Recoil Mass} (Right) of the $Z\rightarrow\mu^+\mu^-$ in the $ZZ$ process using \emph{Kernel Estimation}, with polarization mode $\polLR$. An accuracy of 13~MeV is obtained for the invariant mass measurement, and 28~MeV for the recoil mass measurement. } 
\label{fig:fit_zz}
\end{figure}

The $ZZ$ process is an excellent choice, with the $M_Z$ precisely known to a precision of 2~MeV~\cite{pdg}. The $ZZ$ process has a similar scenario as the Higgs-strahlung process with one $Z$ boson decays to a pair of muons or electrons, and the recoil mass of the Higgs replaced by that of the other $Z$ boson.

By measurement of the invariant mass of the $Z\rightarrow l^+l^-$, the tracking momentum measurement can be calibrated. At $\sqrt{s}=250$~GeV, assuming $M_H=120$~GeV, the $ZZ \rightarrow \mu\mu X/eeX$ has about 40 times larger cross section than that of the $ZH \rightarrow \mu\mu X / eeX$.  With this much larger statistics the $Z$ mass can be measured to a precision of 13 MeV, using channel $ZZ \rightarrow \mu\mu X$, see Figure~\ref{fig:fit_zz}~(Left) for the fit.

The $Z$ recoil mass of the $ZZ$ process can be used to determine and control the center of mass energy and the radiative effects. The Z recoil mass could be determined to a statistical precision of 28 MeV, using channel $ZZ \rightarrow \mu\mu X$, see the fit in Figure~\ref{fig:fit_zz}~(Right).  With this small statistical error, the knowledge of the center of mass energy and the radiative effects could be validated precisely.

On the cross-section measurement, the major systematic error comes from the uncertainty of the efficiency. The procedure to measure the uncertainty of the efficiency is to vary the physics assumptions together with the various background rejection methods to estimate the dependences and covariances between them.  Reminding we applied a very complicated background rejection method. The complicated background rejection mandatorily introduces more difficulties in the measurement of the uncertainty of the efficiency.

For the cross section measurement, we found that the statistical error does not request such a high suppression of the background as for the mass measurement. By removing the Likelihood method and remaining only cuts on some basic variables $M_{dl}$, $P_{Tdl}$, and $P_T$ balance, the statistical error on the cross section measurement only increases by about 10~\% on average. This means a similar statistical error could be obtained with much less sources of systematic error due to the background rejection methods. 

\begin{footnotesize}

\end{footnotesize}

\end{document}